\documentclass[%
reprint,
superscriptaddress,
nofootinbib,
nobibnotes,
amsmath,amssymb,
aps,
prl
]{revtex4-2}

\usepackage[latin1]{inputenc}
\usepackage{graphicx}
\usepackage{dcolumn}
\usepackage{bm}
\usepackage[colorlinks=true,allcolors=blue,bookmarks=true,bookmarksnumbered=true]{hyperref}
\usepackage{multirow}
\usepackage{xcolor}
\usepackage{braket}

\usepackage{natbib}
\usepackage{siunitx}
\usepackage[version=4]{mhchem}
\usepackage{float}
\usepackage{booktabs}

\begin{document}

\title{Evaluation of a \texorpdfstring{\ce{^{88}Sr^+}}{88Sr+} optical clock with a direct measurement of the blackbody radiation shift and determination of the clock frequency}
\author{M. Steinel}
\affiliation{Physikalisch-Technische Bundesanstalt, Bundesallee 100, 38116 Braunschweig, Germany}
\author{H. Shao}
\affiliation{Physikalisch-Technische Bundesanstalt, Bundesallee 100, 38116 Braunschweig, Germany}
\author{M. Filzinger}
\affiliation{Physikalisch-Technische Bundesanstalt, Bundesallee 100, 38116 Braunschweig, Germany}
\author{B. Lipphardt}
\affiliation{Physikalisch-Technische Bundesanstalt, Bundesallee 100, 38116 Braunschweig, Germany}
\author{M. Brinkmann}
\affiliation{Physikalisch-Technische Bundesanstalt, Bundesallee 100, 38116 Braunschweig, Germany}
\author{A. Didier}
\affiliation{Physikalisch-Technische Bundesanstalt, Bundesallee 100, 38116 Braunschweig, Germany}
\author{T. E. Mehlst\"{a}ubler}
\affiliation{Physikalisch-Technische Bundesanstalt, Bundesallee 100, 38116 Braunschweig, Germany}
\affiliation{Leibniz Universit\"{a}t Hannover, Welfengarten 1, 30167 Hannover, Germany}
\author{T. Lindvall}
\affiliation{VTT Technical Research Centre of Finland Ltd, National Metrology Institute VTT MIKES, P.O. Box 1000, 02044 VTT, Finland}
\author{E. Peik}
\affiliation{Physikalisch-Technische Bundesanstalt, Bundesallee 100, 38116 Braunschweig, Germany}
\author{N. Huntemann}
\email[]{nils.huntemann@ptb.de}
\affiliation{Physikalisch-Technische Bundesanstalt, Bundesallee 100, 38116 Braunschweig, Germany}

\date{\today}

\begin{abstract}
We report on an evaluation of an optical clock that uses the $\phantom{}^2S_{1/2} \rightarrow \phantom{}^2D_{5/2}$ transition of a single \ce{^{88}Sr^+} ion as the reference. 
In contrast to previous work, we estimate the effective temperature of the blackbody radiation that shifts the reference transition directly during operation from the corresponding frequency shift and the well-characterized sensitivity to thermal radiation.
We measure the clock output frequency against an independent \ce{^{171}Yb^+} ion clock, based on the $\phantom{}^2S_{1/2} (F=0) \rightarrow \phantom{}^2F_{7/2} (F=3)$ electric octupole (E3) transition, and determine the frequency ratio with a total fractional uncertainty of \num{2.3e-17}.
Relying on a previous measurement of the \ce{^{171}Yb^+} (E3) clock frequency, we find the absolute frequency of the \ce{^{88}Sr^+} clock transition to be $\SI{444779044095485.271 \pm 0.059}{\hertz}$.  
Our result reduces the uncertainty by a factor of $3$ compared to the previously most accurate measurement and may help to resolve so far inconsistent determinations of this value.
We also show that for three simultaneously interrogated \ce{^{88}Sr^+} ions, the increased number causes the expected improvement of the short-term frequency instability of the optical clock without degrading its systematic uncertainty.
\end{abstract}

\maketitle

Optical clocks have demonstrated record fractional uncertainties on the order of and below $10^{-18}$ \cite{bre19, san19}.
In addition to their development as future primary standards of time and frequency, comparisons between optical clocks have been used to provide stringent tests of fundamental physical principles such as the Einstein equivalence principle \cite{lan21,san19} and were employed in searches for ultralight scalar dark matter \cite{rob20,bel20a}.
Optical clocks based on neutral atoms in optical lattices and on ions confined in radio-frequency traps both show significant contributions to their total uncertainty originating from the uncertainty of the Stark shift induced by thermal radiation.
For neutral atoms, the effective temperature of the perturbing field has been determined with high accuracy using resistive temperature sensors at the position of the atoms \cite{blo14} or using temperature controlled environments with high emissivity \cite{bel14}.
For trapped ions, however, the determination is particularly challenging because of the inhomogeneous distribution of heat and the large uncertainty of the emissivity of the components of an ion trap \cite{dol15a}.
Consequently, the uncertainty of the effective temperature of the perturbing thermal field has been on the order of a few Kelvin for many trapped ion systems \cite{dol15a, bre19}.
This technical limitation has been addressed by proper ion trap engineering using materials with a high thermal conductivity.
For such systems, the combination of finite element simulations and measurements using temperature sensors on the trap assembly and infrared cameras enables temperature uncertainties as low as \SI{0.08} K \cite{nor20, nis16}.
In a complementary approach, the in-situ temperature measurement reported in this letter does not require sophisticated techniques but makes direct use of the frequency shift induced by thermal radiation on the  $\phantom{}^2S_{1/2} \rightarrow \phantom{}^2D_{5/2}$ clock transition at \SI{674}{\nano\meter} of a single trapped \ce{^{88}Sr^+} ion. The technique is easily applicable to other ion species like \ce{^{40}Ca^+} \cite{hua19}.

For the demonstration of the technique, we employ a linear ion trap made from printed circuit boards of fiber-reinforced hydrocarbon/ceramics (Rogers 4350B). The low heat conductivity of only \SI{0.69}{\watt\per\meter\per\kelvin}  makes an accurate determination of the effective temperature using well-established techniques challenging. The trap geometry is based on the design discussed in \cite{pyk14}, that provides several trapping regions with 10 individual segments to confine \ce{^{88}Sr^+} ions. We apply radio-frequency and dc voltages to obtain secular trapping frequencies of up to \SI{1320}{\kilo\hertz} and \SI{870}{\kilo\hertz} in the radial and axial direction, respectively. The ion is laser-cooled to \SI{0.62 \pm 0.05}{\milli\kelvin} on the \SI{422}{\nano\meter} electric dipole transition, consistent with the Doppler limitation expected from laser cooling \cite{mad01}. A repump laser at \SI{1092}{\nano\meter} prevents population trapping in the $\phantom{}^2D_{3/2}$ state and is polarization-modulated to enable fast repumping from all $\phantom{}^2D_{3/2}$ Zeeman sublevels. 

During clock operation, the clock transition is periodically interrogated. Each sequence starts with \SI{5}{\milli\second} of laser cooling. State preparation of a single Zeeman sublevel $m_S = -1/2(+1/2)$ of the $^2S_{1/2}$ ground state within \SI{1}{\milli\second} is implemented by continuously driving the \SI{674}{\nano\meter} $m_S = +1/2(-1/2) \rightarrow m_D = -3/2(+3/2)$ transition respectively, together with \SI{1033}{\nano\meter} repumping \cite{roo06}. Mechanical shutters block laser light employed for cooling during the subsequent clock interrogation. Their operation causes a dead time of \SI{5}{\milli\second} at the beginning and the end of the clock interrogation. A successful clock excitation is indicated by absence of fluorescence at the beginning of the subsequent cooling cycle. After \SI{5}{\milli\second} of fluorescence detection, laser light at \SI{1033}{\nano\meter} enables fast depletion of the excited clock state via the $P_{3/2}$ state back to the ground state. 
A valid clock interrogation cycle ends with reappearance of the fluorescence signal within the next cooling period. 

To coherently interrogate the \ce{^{88}Sr^+} clock transition, we set up a laser system at \SI{674}{\nano\meter} which takes advantage of the low frequency instability of a cryogenic silicon resonator \cite{mat17a} at short averaging times and uses an independent \ce{^{171}Yb^+}(E3) clock as the long-term reference~\cite{san19}.  An external-cavity diode laser at \SI{674}{\nano\meter} is stabilized with a bandwidth of \SI{500}{\kilo\hertz} to a piezo-controlled Fabry-Perot resonator with a finesse of $34000$ using a Pound-Drever-Hall locking scheme \cite{dre83}. The transmitted light is frequency-shifted with an acousto-optic modulator (AOM) by about \SI{80}{\mega\hertz}, and used to injection-seed a laser diode. The output power of up to \SI{7}{\milli\watt} is increased  to more than $\SI{50}{\milli\watt}$ using a tapered amplifier. A small fraction of the light is sent to a frequency comb generator and compared to the \ce{^{171}Yb^+}(E3) clock laser light. A discriminator signal is provided using the transfer scheme \cite{tel02b} to steer the AOM and the length of the Fabry-Perot resonator. Apart from a digitally controlled clock offset, the \ce{^{171}Yb^+}(E3) clock laser is stabilized to  the length of the cryogenic resonator via a laser at \SI{1.5}{\micro\meter} making use of the same transfer scheme. An additional AOM provides a digitally-controllable frequency offset of the clock laser light at the position of the \ce{^{88}Sr^+} ion. For all clock laser beams guided in optical fibers, active path-length stabilization is employed.

Neither the \ce{^{88}Sr^+} ground state $S_{1/2}$ nor the excited state $D_{5/2}$ possess sublevels with a magnetic quantum number $m = 0$, consequently all of the possible $(S_{1/2}, m = m_S) \rightarrow (D_{5/2}, m = m_D)$ transition frequencies linearly depend on the magnitude of the magnetic field. Since the corresponding frequency shifts of positive and negative Zeeman sublevels of the ground and excited state with the same absolute magnetic quantum numbers $\lvert m_S \rvert$ and $\lvert m_D \rvert$ average to zero, pairs of transitions are interrogated. To avoid low-frequency magnetic field noise, a single-layer mu-metal shield surrounds the vacuum chamber and low-noise current sources are used to provide a magnetic field of $\approx \SI{4}{\micro\tesla}$. An angle of \SI{30}{\degree} between the propagation direction of the horizontally polarized clock laser light and the applied magnetic field leads to a non-zero excitation probability for all transitions with $\lvert \Delta m \rvert \leq 2$.

In addition to the linear Zeeman shift, the excited state sublevels are affected by tensorial frequency shifts, which scale with $m_D^2$ \cite{dub13}. These are the electric quadrupole shift, which results from a coupling between electric field gradient and electric quadrupole moment of the excited state, and the tensorial part of the Stark shift resulting from the residual rf trapping field. To suppress tensorial shifts, we average multiple pairs of transitions with positive and negative Zeeman shift and with different $m_D^2$. The average frequency of transitions to all excited state sublevels is intrinsically free of tensor shifts \cite{dub05}. In first-order perturbation theory, the same frequency is obtained by averaging only two pairs of transitions using appropriate weights \cite{dub13}. Since magnetic field noise determines the maximum coherent interrogation time for our setup, we use a numerical optimization to minimize the frequency instability of the clock under the condition of tensor shift cancellation using individually adjusted interrogation pulse durations on each transition pair. We find the lowest frequency instability if the $\ket{S, \pm 1/2} \rightarrow \ket{D, \pm 3/2}$ transition pair is interrogated twice as often as the  $\ket{S, \pm 1/2} \rightarrow \ket{D, \pm 5/2}$ transition pair, with pulse durations of \SI{87.5}{\milli\second} and \SI{35}{\milli\second} and averaged with weights of $5/6$ and $1/6$, respectively \cite{supp}. 

The center frequency of each transition involved in the averaging scheme discussed above is tracked individually by an integrating servo system and averaged in post-processing to obtain the frequency difference between the clock laser and the unperturbed $S_{1/2} \rightarrow D_{5/2}$ transition. Each transition is interrogated with single rectangular clock laser pulses with positive ($+$) and negative ($-$) detunings $\Delta\nu$ from the center frequency, such that the expected excitation probability is \SI{50}{\percent} of the maximum. The corresponding servo systems shift the center frequencies by $0.15 (n_+ - n_-) \times \Delta\nu$ with $n_{+/-}$ being the number of successful excitations after four interrogations \cite{pei06}. 

\begin{figure}[t]
    \centering
    \includegraphics[width=0.85\linewidth]{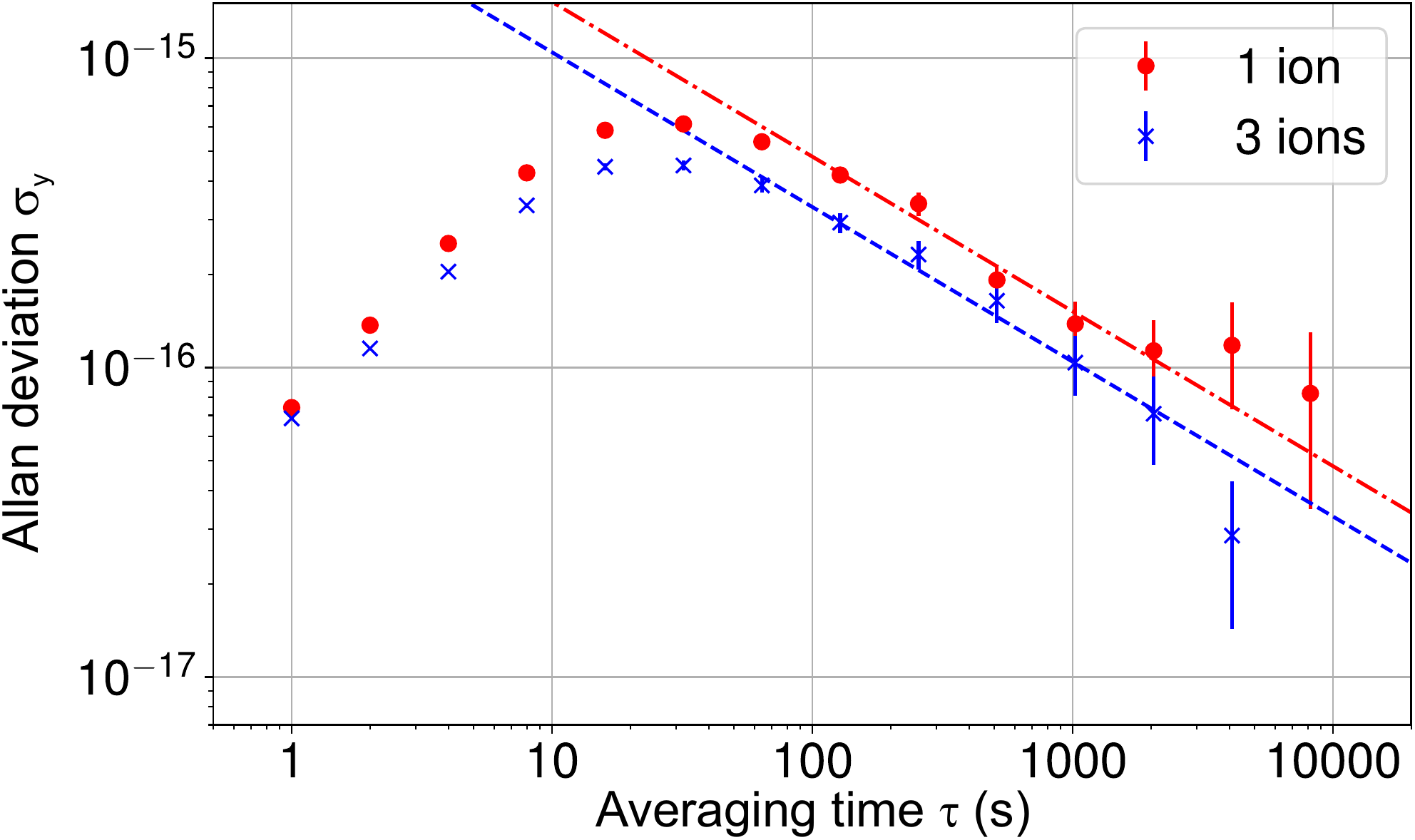}
    \caption{Fractional instability determined by the Allan deviation $\sigma_y$ of the frequency ratio $\nu_{\ce{^{88}Sr^+}} / \nu_{\ce{^{171}Yb^+}}$ for a single measurement run. The long-term behaviour predicted by a Monte-Carlo simulation of the servo algorithm assuming quantum projection noise only is indicated by the dashed-dotted line. Also shown is the result of the comparison when the \ce{^{88}Sr^+} clock is operated with three ions. The corresponding simulation considers the increased detection and cooling time (dashed line).}
    \label{fig:AllanDeviation}
\end{figure}

To experimentally investigate the frequency instability of the \ce{^{88}Sr^+} clock, we compare it to the \ce{^{171}Yb^+}(E3) clock \cite{san19}, which employs the electric octupole (E3) transition at \SI{467}{\nano\meter}. For the latter, a frequency instability of $\num{1e-15}(\tau /\si{\second})^{-1/2}$ has recently been reported in Ref.~\cite{doe21}. Fig.~\ref{fig:AllanDeviation} shows the fractional Allan deviation $\sigma_y$ of the frequency ratio $\mathfrak{R} = \nu_{\ce{^{88}Sr^+}} / \nu_{\ce{^{171}Yb^+}}$ for a typical measurement run recorded from MJD 59627 to MJD 59634 (17 Feb 2022 to 24 Feb 2022). A Monte-Carlo simulation of the servo algorithm assuming only quantum projection noise predicts an instability $\sigma_y = \num{4.8e-15}(\tau /\si{\second})^{-1/2}$, with which the experimental data is in good agreement.

The frequency instability is expected to decrease with $1/\sqrt{N}$ where $N$ is the number of simultaneously interrogated ions. Consequently, the measurement time required to reach a given statistical uncertainty is reduced by $N$, if the experimental sequence remains otherwise unchanged. This particular advantage of a so-called multi-ion clock \cite{her12b} can be exploited with ion species that show a  small sensitivity to electric field gradients, such as \ce{^{115}In^+} \cite{her12b}, \ce{^{27}Al^+} \cite{bre19}, \ce{Sn^{2+}} \cite{saf22}, \ce{Pb^{2+}} \cite{bel20b}, or other  atomic species when multi-ion related shifts are zeroed by dynamic decoupling or state averaging \cite{arn16, tan19}. In these cases, fractional frequency uncertainties of $10^{-19}$ and below can be achieved \cite{kel19}. To verify the fundamental $1/\sqrt{N}$ scaling for our system, we also operate the clock with three \ce{^{88}Sr^+} ions and align the magnetic field at an angle of \SI{54.7 \pm 0.4 }{\degree} to the trap axis to minimize the frequency shift induced by the electric field gradient along the trap axis \cite{tan19}. The corresponding frequency instability follows $\sigma_y = \num{3.3e-15}(\tau /\si{\second})^{-1/2}$ as shown in Fig.~\ref{fig:AllanDeviation}. The instability does not decrease by the expected factor $\sqrt{3}$ compared to the single ion result, because the detection and cooling times were increased to \SI{15}{\milli\second} each to reliably determine the number of excited ions from the integrated single photon detector counts of all ions. The mean frequency ratio determined with three ions agrees with the single ion result discussed below up to a fractional difference of $2.7(3.3) \times 10^{-17}$.

\begin{table}[t]
    \centering
    \begin{tabular}{l c c}
        \toprule 
        \toprule 
        Shift effect & $\delta\nu/\nu_{0}$ ($10^{-18}$) & $u(\delta\nu)/\nu_{0}$ ($10^{-18}$)  \\
        \midrule
        Blackbody radiation & 537.9 & 7.6 \\
        Micromotion &  -15.1 & 6.6 \\
        Collisions & 0.0 & 0.5 \\
        Thermal motion &  -0.932 & 0.093 \\
        Quadratic Zeeman & 0.1409 & 0.0003 \\
        \midrule
        Total & 522 & 10 \\
        \bottomrule
        \bottomrule
    \end{tabular}
    \caption{Fractional frequency shifts $\delta \nu/\nu_0$ and corresponding uncertainties $u(\delta\nu)/\nu_0$ considered for the realization of the unperturbed $S_{1/2} \rightarrow D_{5/2}$ transition frequency $\nu_0$ of the \ce{^{88}Sr^+} single ion clock for low rf power of the ion trap.}
    \label{tab:SysShifts}
\end{table}

\begin{figure}[t]
    \centering
    \includegraphics[width=0.85\linewidth]{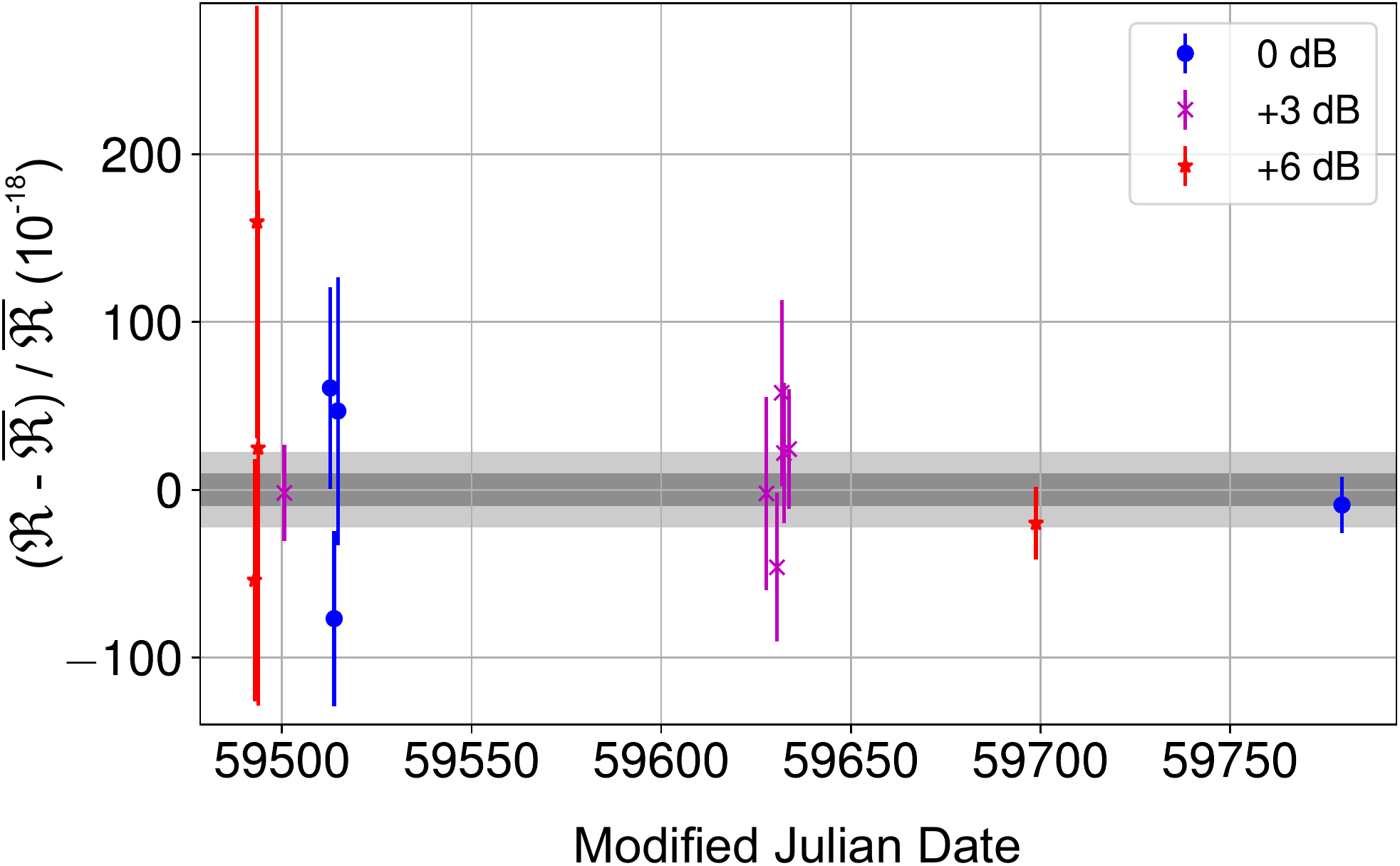}
    \caption{Measurements of the ratio $\mathfrak{R}$ of the \ce{^{88}Sr^+} and \ce{^{171}Yb^+}(E3) clock frequencies for different settings of the relative ion trap drive powers, distributed over 8 months.  The error bars indicate the statistical uncertainty. All data are consistent with the weighted mean value $\overline{\mathfrak{R}}$. The grey shaded area shows its statistical uncertainty of \num{1.0e-17}, and the light grey area corresponds to the total uncertainty (\num{2.3e-17}).}
    \label{fig:SrYbRatio}
\end{figure}

Shifts considered in the evaluation of the \ce{^{88}Sr^+} single-ion clock are summarized in Tab.~\ref{tab:SysShifts}. While the dominant shift resulting from thermal radiation is investigated using a novel approach discussed below, the evaluation of all other shift effects follows well-established methods \cite{dub13}. The second largest shift results from excess micromotion causing Stark and second-order Doppler shifts. The corresponding rf field is determined using the photon correlation technique \cite{kel15} with a minimum amplitude of \SI{320 \pm 70}{\volt\per\meter}. It is mainly aligned along the trap axis which prevents a compensation using static electric fields. Because of the negative differential polarizability $\Delta \alpha = \SI{-4.7938 \pm 0.0071e-40}{\joule\meter^2\per\volt^2}$, the second order Doppler and Stark shift cancel out at the so-called magic rf frequency of about \SI{14.4}{\mega\hertz} \cite{dub14}. For the rf trap drive frequencies of \SI{12.8}{\mega\hertz} and \SI{13.28125}{\mega\hertz} used for the measurements reported here, the relative difference of the absolute values of both shifts are 0.24 and 0.17, respectively. Excess micromotion is minimized and determined before and after each clock run along three non-coplanar orientations. Significant deviations from a constant amplitude over each measurement run have not been observed. 

The frequency shifts from thermal ion motion are calculated using the ion temperature inferred from measurements of the relative excitation probability on the carrier and red sideband transitions. For the axial and radial directions at secular frequencies of \SI{870}{\kilo\hertz} and \SI{1320}{\kilo\hertz}, we find ion temperatures of \SI{0.534 \pm 0.097}{\milli\kelvin} and \SI{2.02 \pm 0.24}{\milli\kelvin}, respectively. These measurements include the temperature rise during clock interrogation resulting from motional heating of the trapped ion with $\mathrm{d}T_{\text{ax}}/\mathrm{d}t = \SI{0.73 \pm 0.026 }{\milli\kelvin\per\second}$ and $\mathrm{d}T_{\text{rad}}/\mathrm{d}t = 16.6(2.2)\si{\milli\kelvin\per\second}$. While the linear Zeeman shifts average to zero for the clock output signal, they can be employed to determine the quadratic Zeeman shifts on the fly using the sensitivity coefficient given in Ref.~\cite{dub13}. During the clock interrogation, mechanical shutters block all laser beams except the \SI{674}{\nano\meter} clock laser. 

We estimate the frequency shift induced by collisions of the trapped ion with background gas molecules from position changes of a two-ion crystal \cite{han19a}. Our measurements were performed using \ce{^{171}Yb^+} and \ce{^{174}Yb^+} in the same ion trap segment and indicate a maximum fractional frequency shift of \num{5e-19} which we consider as the corresponding uncertainty.

\begin{figure}[t]
    \centering
    \includegraphics[width=0.85\linewidth]{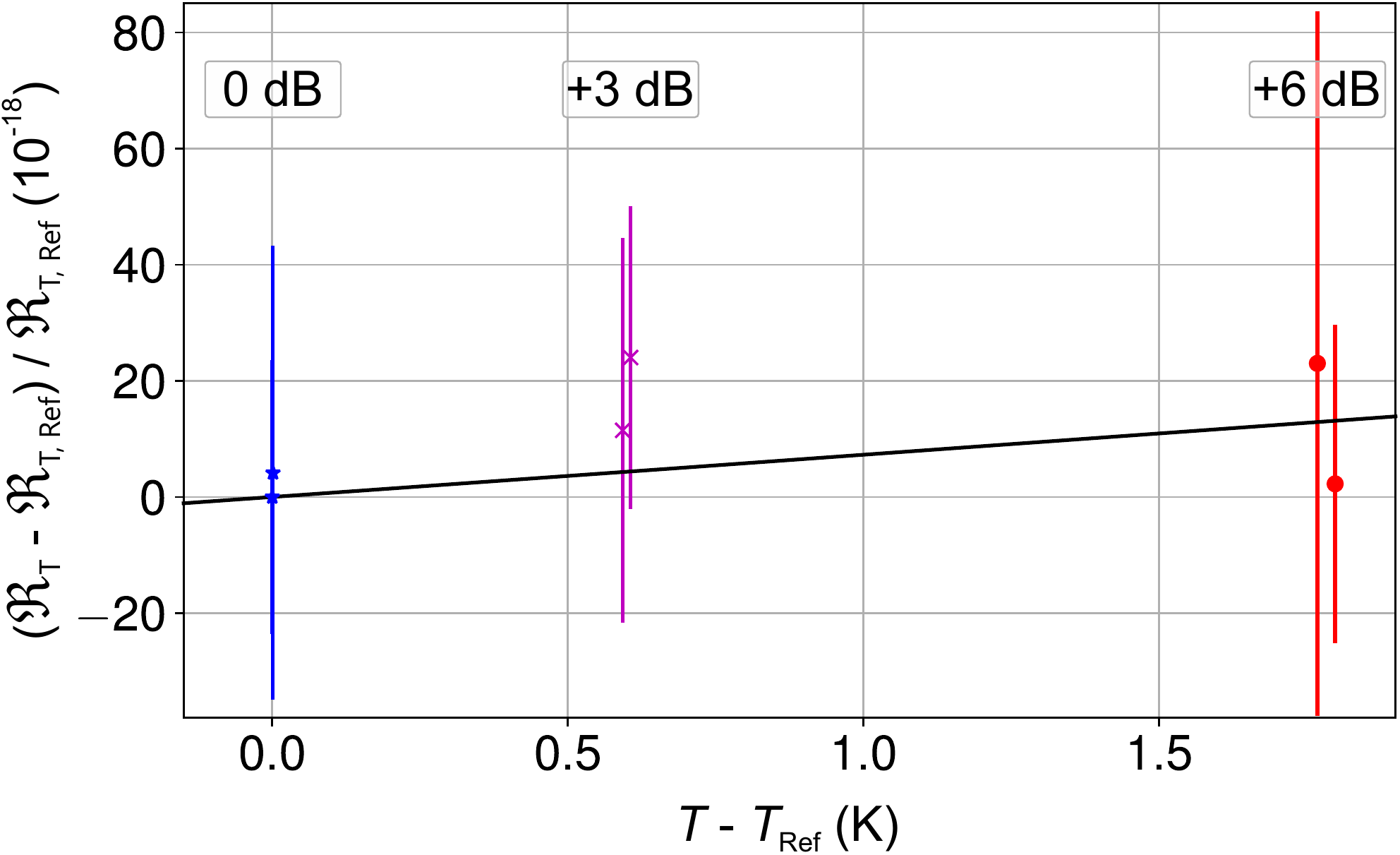}
    \caption{Relative difference of the frequency ratio $\mathfrak{R}_T$ of the \ce{^{88}Sr^+} clock and an independent fully corrected \ce{^{171}Yb^+}(E3) clock from the measurement with the smallest statistical uncertainty $\mathfrak{R}_{T,\text{Ref}}$. For the \ce{^{88}Sr^+} clock all frequency shifts are corrected except for the BBR shift. The ion trap of the \ce{^{88}Sr^+} clock is operated with different settings of the relative rf drive power $P/P_0$, indicated by symbol and color. Changes in $T_{\text{BBR}}$ resulting from variations in the ambient temperature $T_B$ for measurements with the same rf power have been corrected for. The BBR shift obtained via least-squares regression is indicated by the black line. The error bars indicate only statistical uncertainties.}
    \label{fig:BBRRatio}
\end{figure}

The largest deviation between the observed and the unperturbed clock transition frequency is caused by the quadratic Stark shift of thermal radiation, typically referred to as the blackbody radiation (BBR) shift 
\begin{equation}
    \Delta \nu_{\text{BBR}} = - \frac{1}{2 h} \langle E^2  (T_{\text{BBR}}) \rangle \Delta \alpha_0 (1 + \eta (T_{\text{BBR}})),
    \label{eq:bbrgen}
\end{equation}
where $h$ is the Planck constant, $\langle E^2  (T_{\text{BBR}}) \rangle$ is the mean squared electric field of a black body at temperature $T_{\text{BBR}}$, $\Delta \alpha_0$ is the static differential polarizability and $\eta (T_{\text{BBR}})$ corrects for the wavelength dependence of the differential polarizability over the BBR spectrum \cite{jia09}. 

The temperature $T_{\text{BBR}}$ can be described by the background temperature $T_B$ measured outside the vacuum chamber with accurate temperature sensors and the temperature rise $\Delta T$ resulting from radio-frequency losses that heat the ion trap assembly during operation. 
This rise of the effective temperature $\Delta T$ is usually determined by finite-element method modelling, complemented with infrared camera observations and sensor measurements \cite{dol15a}. 
While infrared camera observations require infrared-transparent windows, sensor measurements are only reliable when the trap is made from materials with high thermal conductivity and direct rf-induced heating of the sensors is avoided. 
These requirements are usually only fulfilled in highly specialized ion traps \cite{nor20, nis16}, but here we show that $\Delta T$ can also be measured spectroscopically. 
The spectroscopic determination is particularly important for setups where the heat is predominantly generated inside trap assembly so that it can not be easily measured outside the vacuum chamber. 
For moderate temperature increases of the ion trap and negligible heat dissipation via radiation, $\Delta T$ is proportional to the power provided by the RF drive $P_{\text{rf}}$.
Under these constraints, frequency measurements at different $P_{\text{rf}}$ of optical transitions with large, but known BBR shift sensitivity enable extrapolation to $\Delta T = 0$.

For \ce{^{88}Sr^+}, the sensitivity of the clock transition frequency to thermal radiation can be well approximated by $\Delta \nu_{\text{BBR}}(T) = \SI{3.0616 \pm 0.0046e-11}{\hertz\per\kelvin^4}\times T^4$ at room temperature \cite{dub14}. Fig.~\ref{fig:SrYbRatio} shows measurements of the frequency ratio $\mathfrak{R}$ for different settings $P/P_0$ of the rf drive power of the \ce{^{88}Sr^+} ion trap. The temperature rise of $\Delta T = 0.6(1.0)\si{\kelvin}$ at $P_0$ is derived from a least-squares regression of $\mathfrak{R}_{T}(T_B, P/P_0) = \Delta \nu_{\text{BBR}}(T_B + P/P_0 \times \Delta T)/\nu_{\ce{^{171}Yb^+}}$. To improve visibility of the relative change of $\mathfrak{R}_T$ in Fig.~\ref{fig:BBRRatio}, data collected for the same $T_B$ and $P/P_0$ setting are averaged.

Using all recorded data, the frequency ratio of the unperturbed \ce{^{88}Sr^+} and $\phantom{}^2S_{1/2} (F=0) \rightarrow \phantom{}^2F_{7/2} (F=3)$ \ce{^{171}Yb^+}(E3) clock transitions is $\mathfrak{R} = \num{0.692 671 163 215 966 050 \pm 0.000 000 000 000 000 016}$. The total relative uncertainty of \num{2.3e-17} results from the statistical uncertainty (\num{1.0e-17}) and the systematic uncertainties of the \ce{^{88}Sr^+} and \ce{^{171}Yb^+}(E3) clocks of \num{2.0e-17} and \num{2.7e-18}\cite{san19}, respectively.

\begin{figure}[t]
    \centering
    \includegraphics[width=0.85\linewidth]{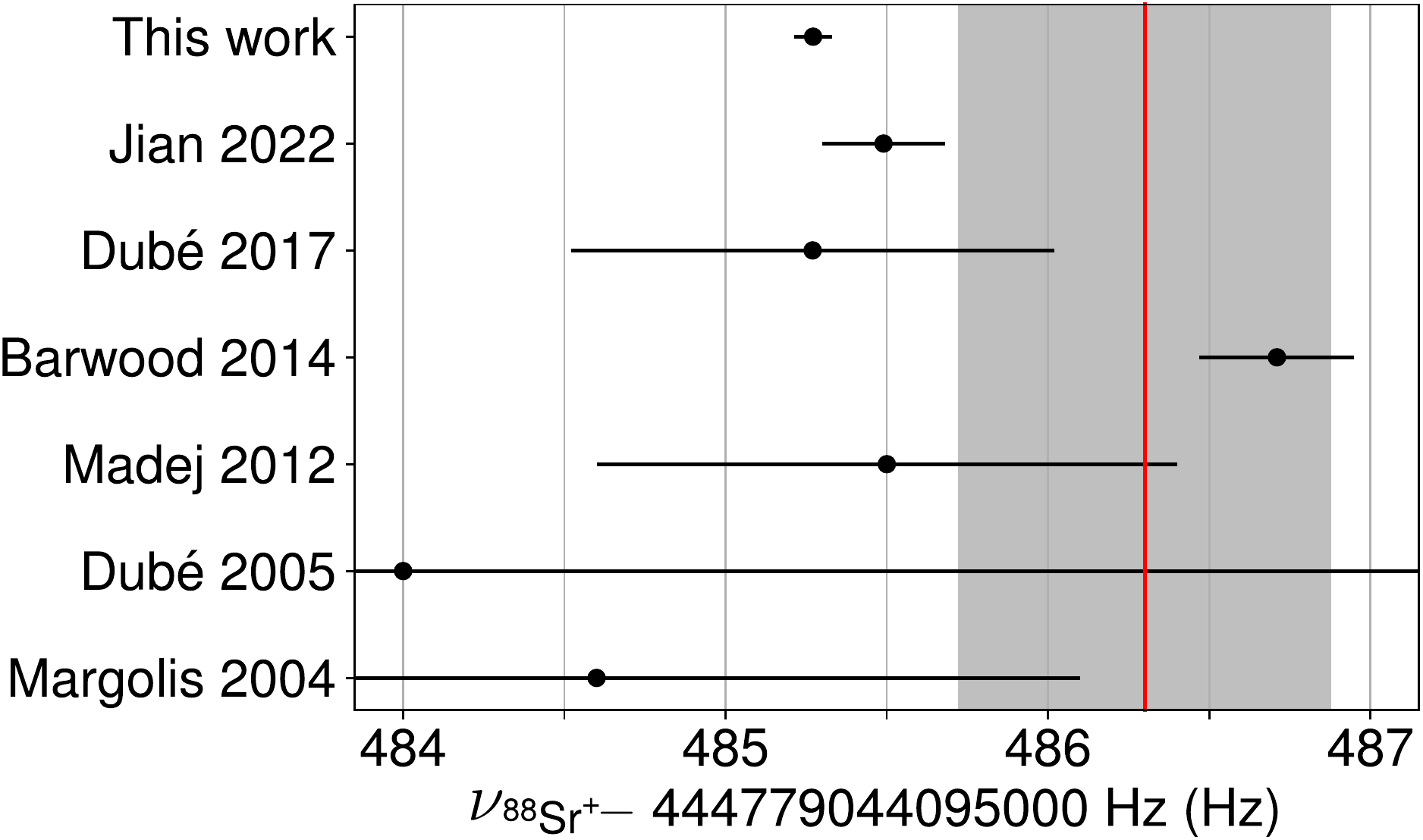}
    \caption{Results of measurements of the  \ce{^{88}Sr^+} clock transition frequency obtained in previous \cite{mar04, dub05, mad12, bar14b, dub17, jia22} and this work. The recommended value for $\nu_{\ce{^{88}Sr^+}}$ as a secondary representation of the second (SRS) approved in 2021 \cite{cct21} is shown by the red line and the shaded area is its uncertainty.} 
    \label{fig:SrFreqTableRatio}
\end{figure}

We calculate the absolute frequency of the \ce{^{88}Sr^+} clock transition from $\mathfrak{R}$ and an accurate measurement of the \ce{^{171}Yb^+}(E3) clock transition frequency previously performed in our laboratory \cite{lan21} to be $\nu_{\ce{^{88}Sr^+}} = \SI{444779044095485.271 \pm 0.059}{\hertz}$. This result is compared with previously published values \cite{mar04, dub05, mad12, bar14b, dub17, jia22} and the recommended value of the standard frequency \cite{cct21} in Fig.~\ref{fig:SrFreqTableRatio}. The previous two most accurate determinations show a discrepancy of $4 \sigma$. Our result agrees with all except the determination from Ref.~\cite{bar14b} and we improve the total uncertainty compared to the previous most accurate measurement \cite{jia22} by a factor of $3$. This information is particularly relevant for the next adjustment of the recommended value of the standard frequency.

In summary, we have operated an optical clock based on the $\phantom{}^2S_{1/2} \rightarrow \phantom{}^2D_{5/2}$  transition of a single trapped \ce{^{88}Sr^+} ion to determine its frequency against an independent \ce{^{171}Yb^+}(E3) clock. The frequency ratio with a fractional uncertainty of \num{2.3e-17} is among the most precisely measured natural constants to date and the inferred absolute frequency will support an improved recommended value. This result is particularly important for the practical realization of the System International base unit Hertz with \ce{^{88}Sr^+} ion clocks \cite{cct21}. We have also shown that the clock can be operated with multiple ions to reduce the frequency instability. Our method to determine the effective temperature is directly applicable for various ion species and provides a similar uncertainty like more involved techniques \cite{dol15a}. It is particularly advantageous for clock systems with ancillary transitions that feature a large but known differential polarizability such as \ce{^{27}Al^+}/\ce{^{40}Ca^+} \cite{han19, hua19}, \ce{^{176}Lu^+} \cite{arn18} and \ce{^{115}In^+}/\ce{^{172}Yb^+} \cite{kel19a}. 

\begin{acknowledgements}
We acknowledge support by the projects 20FUN01 TSCAC and 17FUN07 CC4C, which have received funding from the EMPIR programme co-financed by the Participating States and from the European Union's Horizon 2020 research and innovation programme, and by the Deutsche Forschungsgemeinschaft (DFG, German Research Foundation) under SFB~1227 DQ-\textit{mat} -- Project-ID 274200144 -- within project B02.
This work was partially supported by the Max Planck--RIKEN--PTB Center for Time, Constants and Fundamental Symmetries.
\end{acknowledgements}

\section{Supplemental Material}
The frequency $\nu^{m_S}_{m_D}$ of each $\ket{S, m_S} \rightarrow \ket{D, m_D}$  transition of \ce{^{88}Sr^+} linearly depends on the magnetic field and $g_D m_D - g_S m_S$, where $g_D$ and $g_S$ are the Land\'{e} $g$-factors. The corresponding frequency shift is avoided during clock operation by averaging transitions with positive and negative Zeeman shifts. In addition, tensorial shifts that show a dependence on $m_D^2$ need to be suppressed. For \ce{^{88}Sr^+} this is typically done by averaging the transition frequencies to all excited state sublevels \cite{dub05}. To first order of perturbation theory, the same frequency is obtained by averaging only two pairs of transitions using appropriate weights $w_{m_D}$ \cite{dub13}
\begin{equation}
    \overline{\nu} = \frac{\sum_{m_D}{{ w_{m_D} \overline{\nu}_{m_D}}}}{\sum_{m_D}{w_{m_D}}},\ \overline{\nu}_{m_D} = \frac{\nu^{+1/2}_{+m_D} + \nu^{-1/2}_{-m_D}}{2}.
\end{equation}

In our experiment the maximum coherent interrogation time is limited by magnetic field noise, and is therefore inversely proportional to $g_D-m_D - g_Sm_S$ for the different available transitions pairs. Interrogating the separate Zeeman components with adapted interrogation times and repetition rates can improve the clock instability in comparison to an interrogation sequence which applies the same parameters for each pair. This procedure is adequate when the instability is predominantly dominated by quantum projection noise and tensorial shifts vary only slowly with respect to the interrogation cycle.

The variance of the mean transition frequency $\sigma^2_{\overline{\nu}}$ is given by
\begin{equation}
    \sigma^2_{\overline{\nu}} = \frac{\sum_{m_D}{{ w_{m_D}^2 \sigma^2_{\overline{\nu}_{m_D}}}}}{\left( \sum_{m_D}{w_{m_D}} \right)^2}
\end{equation}
with the individual variances
\begin{equation}
    \sigma^2_{\overline{\nu}_{m_D}} =\frac{\left( a\Delta \nu_{m_D} \right)^2}{n_{m_D}}
\end{equation}
with the linewidth of the transition $\Delta \nu_{m_D}$, the number of interrogations of a transition during a fixed time interval $n_{m_D}$ and a common proportionality factor $a$ between the variance and linewidth. 

We minimize $\sigma^2_{\overline{\nu}}$ for integer $n_{m_D}$, under the constraint that $n = \sum_{m_D}{n_{m_D}}$ is fixed and the weights $w_{m_D}$, including $w_{m_D} = 0$, preserve tensorial shift cancellation. We find that the transitions $\ket{S, \pm 1/2} \rightarrow \ket{D, \pm 3/2}$ and $\ket{S, \pm 1/2} \rightarrow \ket{D, \pm 5/2}$ with the  weights of $w_{3/2} = 5/6,\, w_{5/2} = 1/6$ and number of interrogations $n_{3/2} = 2,\, n_{5/2} = 1$ gives the lowest instability.

\end{document}